\documentclass[]{spie}  

\usepackage{amsmath,amsfonts,amssymb}
\usepackage{graphicx}
\usepackage{subfigure}
\usepackage[colorlinks=true, allcolors=blue]{hyperref}

\newcommand\numberthis{\addtocounter{equation}{1}\tag{\theequation}}

\newcommand{\um}{\rm\thinspace $\mu$m}
\newcommand{\ms}{\rm\thinspace ms}

\newcommand{\keV}{\rm\thinspace keV}
\newcommand{\eV}{\rm\thinspace eV}

\title{Augmenting astronomical X-ray detectors with AI for enhanced sensitivity and reduced background}

\author[a]{D.~R.~Wilkins}
\author[a]{A.~Poliszczuk}
\author[b]{B.~Schneider}
\author[b]{E.~D.~Miller}
\author[a]{S.~W.~Allen}
\author[b]{M.~Bautz}
\author[a]{T.~Chattopadhyay}
\author[c]{A.~D.~Falcone}
\author[b]{R.~Foster}
\author[b]{C.~E.~Grant}
\author[a]{S.~Herrmann}
\author[d]{R.~Kraft}
\author[a]{R.~G.~Morris}
\author[d]{P.~Nulsen}
\author[a]{P.~Orel}
\author[d]{G.~Schellenberger}
\affil[a]{Kavli Institute for Particle Astrophysics and Cosmology, Stanford University, 452 Lomita Mall, Stanford, CA 94305, USA}
\affil[b]{MIT Kavli Institute for Astrophysics and Space Research, 77 Massachusetts Avenue, Cambridge, MA 02139, USA}
\affil[c]{Pennsylvania State University, Department of Astronomy and Astrophysics, University Park, Pennsylvania, United States}
\affil[d]{Harvard–Smithsonian Center for Astrophysics, 60 Garden Street, Cambridge, MA 02138, USA}

\authorinfo{Correspondence e-mail: dan.wilkins@stanford.edu}

\begin{document} 
\maketitle

\begin{abstract}
Bringing artificial intelligence (AI) alongside next-generation X-ray imaging detectors, including CCDs and DEPFET sensors, enhances their sensitivity to achieve many of the flagship science cases targeted by future X-ray observatories, based upon low surface brightness and high redshift sources. Machine learning algorithms operating on the raw frame-level data provide enhanced identification of background vs. astrophysical X-ray events, by considering all of the signals in the context within which they appear within each frame. We have developed prototype machine learning algorithms to identify valid X-ray and cosmic-ray induced background events, trained and tested upon a suite of realistic end-to-end simulations that trace the interaction of cosmic ray particles and their secondaries through the spacecraft and detector. These algorithms demonstrate that AI can reduce the unrejected instrumental background by up to 41.5 per cent compared with traditional filtering methods. Alongside AI algorithms to reduce the instrumental background, next-generation event reconstruction methods, based upon fitting physically-motivated Gaussian models of the charge clouds produced by events within the detector, promise increased accuracy and spectral resolution of the lowest energy photon events.
\end{abstract}


\section{Introduction}
\label{sec:intro}  
Sensitive X-ray imaging detectors will be a cornerstone of the next generation space-based X-ray observatories, including the next European-led flagship X-ray mission \textit{Athena}\cite{athena}, proposed NASA probe-class missions including the \textit{Advanced X-ray Imaging Satellite, AXIS} \cite{axis}, the \textit{High Energy X-ray Probe, HEX-P} \cite{hexp} and \textit{ARCUS}\cite{arcus}, as well as a future US-led X-ray flagship recommended by the \textit{2020 Decadal Survey of Astronomy and Astrophysics}\cite{decadal}, akin to the \textit{Lynx} mission concept\,\cite{lynx}. Many of the flagship science cases for these missions\cite{athena_science} rely on high-sensitivity imaging of low surface brightness X-ray sources at moderate to high redshifts. These include the proposed deep X-ray surveys that will find the seeds of the supermassive black holes in the early Universe \cite{lynx_science}, and measurements of the thermodynamics and elemental abundances in the outskirts of galaxies and galaxy clusters, that represent a key tracer of structure formation, AGN feedback and the cycling of baryons within galactic environments.

Recent technological advances have paved the way for the advancement of high-frame rate, low-read noise X-ray imaging detectors \cite{xray_speed}, notably the DEPFET detector array at the heart of the \textit{Athena Wide Field Imager} (WFI) \cite{wfi}, and next-generation CCD detectors that will form the focal plane of the \textit{AXIS}\cite{axis_focal_plane}.

The sensitivity of these instruments, however, is limited by the instrumental background, which over the 2-10\keV\ band-pass is dominated by signals induced by charged cosmic ray particles interacting with the spacecraft and the detector. Energy deposited in the detector by these particles, as well as the secondaries they produce as they traverse the spacecraft, induces signals or `events' that can be confused with astrophysical X-rays that arrive at the detector via the telescope optics. While 98 per cent of these background events can be filtered out of the final data that are used for the science analysis using traditional methods, the remaining unrejected background, comprised of so-called `valid events' can still represent a significant signal that limits the sensitivity of the detector to the faintest X-ray sources.  Traditional filtering methods operate on the basis of their energy (charged particles often deposit more energy than can be ascribed to a single photon that could have been focused by the mirrors) or the pattern of charge deposited across the pixels of the detectors (while X-ray signals are limited to discrete islands of just a few pixels, charged particles will often leave extended tracks as they pass through the detector).

A further limitation of these detectors is their ability to detect the lowest energy X-ray photons, which become increasingly important when observing celestial sources with soft X-ray spectra, including stars, the circumgalactic medium (CGM) of individual galaxies, the warm-hot component of the intergalactic medium (WHIM), groups of galaxies, and high-redshift galaxy clusters and AN. When a photon is absorbed in the silicon detector, a cloud of electrons is produced that drifts through an electric field to the collection gates for each pixel through which the corresponding signal is read out (hereafter referred to as \textit{readout gates}). During this drift, the electron cloud diffuses, and depending how close to the pixel boundary the photon was absorbed, a significant fraction of these electrons end up being recorded in the neighboring pixel, resulting in so-called `split events'. This phenomenon is well-understood in CCDs and other pixelated silicon X-ray detectors. Single and double pixel events are typically considered to be valid X-rays, in addition to triple and quadruple pixel events, which are more common in those detectors with smaller pixel sizes. The total energy of an event is reconstructed by summing the signal in pixels that exceeds some threshold (referred to as the split threshold, designed to filter out spurious signals from electronic read noise in the empty pixels). In back-illuminated devices, the lowest energy photons will be absorbed in a relatively shallow depth of silicon, furthest from the readout gates; thus the electrons from the lowest energy events will diffuse the furthest. This can result in a significant fraction of the energy of the low-energy photon events being carried by electrons into the outer pixels of the event that never rise above the split threshold. The signal in pixels below the split threshold is lost, leading to an underestimate of the photon energy and a reduction in the sensitivity of the device to the lowest energy photons.

These proceedings report on an ongoing program to develop enhanced methods to filter the cosmic ray-induced background and to improve the energy reconstruction of photon events for next-generation X-ray imaging detectors. These methods are based upon machine learning or artificial intelligence (AI) algorithms as well as statistical modeling of raw event data from the detector, built upon a physical understanding of how charged particles and photons interact with the detector.

\section{Cosmic ray interactions with the spacecraft and detector}
We are developing AI/ML algorithms to detect and filter the component of the instrumental background that is induced by cosmic rays. To this end, we have developed an end-to-end pipeline to simulate realistic detector frames containing both cosmic ray-induced background events, and genuine astrophysical X-ray events that include a realistic description of interactions between cosmic ray particles and the spacecraft and detector, as well as realistic signal generation from these events within the detector.

Interactions between the cosmic ray particles and the spacecraft/detector system were simulated using the Monte Carlo particle interaction code, \textsc{geant4}\cite{geant4}, specifically using a mass model of the system developed to simulate the \textit{Athena Wide Field Imager} \cite{grant+2020,miller_jatis} and a \textit{space physics} physics list to model the relevant particle interactions. This work has been conducted as part of the AREMBES project\cite{wfi_bkg}, which aims to study and mitigate the background on the \textit{Athena} X-ray observatory. While many of the effects observed in the \textit{Athena} simulations are representative of many future mission concepts, alternative mass models can be specialised to specific mission designs.

These simulations trace primary cosmic ray protons through the mass model of the spacecraft and instrument. At each step of the simulation, the primary protons may interact with materials in the mass model and produce one or more secondary particles, including further protons, electrons, positrons and X-ray or $\gamma$-ray photons, including a continuum of X-ray emission in addition to specific fluorescent lines from the materials present (most notably the Al K$\alpha$ fluorescence line at 1.5\keV). The detector itself is defined within the mass model as a sheet of silicon. As any of the primary or secondary particles traverse the detector, the energy they deposit is recorded on a 1\um\ resolution three-dimensional grid. The energy of the primary protons is sufficiently high that they are considered to be minimum ionising particles (MIPs), meaning that they are not stopped in the silicon `detector' but rather deposit energy as they pass through, while secondary photons are absorbed at a point within the silicon at a depth that depends upon their energy.

We assume that the detector is read out frame by frame, following some integration time (nominally 5\ms\ per frame for the \textit{Athena} WFI), the signal recorded in each pixel (\textit{i.e.} the number of electrons liberated within the silicon) is directly proportional to the energy deposited in that pixel during that time. We simulate frames that would be read out from each $512 \times 512$ pixel quadrant of the WFI detector.

\section{Charge diffusion and signal generation within the detector}
Once energy has been deposited at a point within the silicon detector, a cloud of $N$ electrons is generated, where $N$ is drawn from a Gaussian distribution with a mean of one electron per 3.65\eV\ of energy deposited, and variance determined by the Fano factor (the ratio of the variance to the mean) for Silicon, $F = 0.12$. These electrons must then be tracked as they drift from the deposition site to the location of the readout gates to predict the signal that is recorded in each pixel (\textit{i.e.} the number of electrons that reaches each gate)\cite{miller+2022}.

In addition to cosmic ray-induced signals, simulated X-ray signals are added to each frame, depositing the energy from each photon at a single location within the silicon. The $(x,y)$ positions of each photon are drawn from uniform distributions (\textit{i.e.} modelling a uniform source of emission) and the energy is drawn from a distribution representing the spectrum of the source. The simplest cases are a uniform spectral distribution, or a log-uniform distribution, representing a power law with photon index $\Gamma = 1$. The latter case is adopted here, however we aim to develop background filtering algorithms that are not sensitive to the specific spectrum of the source being observed. The depth at which the photon is absorbed is drawn from an exponential distribution characterised by the energy-dependent attenuation length of X-ray photons in silicon. In general, the attenuation length increases monotonically between 0.1\keV\ and the silicon K absorption edge at 1.8\keV, where the attenuation length drops sharply due to enhanced photoelectric absorption, before rising again as the energy increases\cite{burke+1997}.

In this work, we follow the electrons from the energy deposition site to the readout gates using two methods. The first method transports the electron cloud in a simulation of the electric field through the device, using the \textsc{poisson\_ccd} code\cite{miller+2022,lamarr+2022}. \textsc{poisson\_ccd}\cite{poissonccd} traces the path of each electron, including both the drift induced by the electric field, and random diffusion, and brings each electron to the specific readout gate for each pixel.

The second method employs an analytic model for the diffused radius of the charge cloud\cite{poliszczuk+2023}. If energy deposition, $E$, by a photon or charged particle liberates a cloud of electrons within initial radius $\sigma_0$ at height $z$ above the readout gates, the diffused cloud at the readout plane is modeled by a two-dimensional Gaussian with width defined by the standard deviation, $\sigma$: \cite{iniewski+2007,veale+2014}
\begin{equation}
    \sigma(E, z) = \sigma_0(E) + 1.15\,f_\mathrm{diff}\,z\sqrt{\frac{2k_\mathrm{B}T}{eV}}
\end{equation}
$k_B$ is the Boltzmann constant, $T$ is the operating temperature of the detector, determining the magnitude of the diffusion term for the electron cloud, $e$ is the charge of an electron, and $V$ is the detector bias voltage that produces the electric field causing the electrons to drift towards the readout gates. $f_\mathrm{diff}$ is a multiplicative constant, known as the diffusion multiplier, that is applied to the diffusion term at each step in the \textsc{poisson\_ccd} simulation, and to the final cloud radius in the analytic model. The diffusion multiplier can be understood in the context of the \textit{effective mass}, $m_*$ of electrons within the band structure of the silicon, and the precise value depends on the specific properties of each device and the silicon wafer from which it is manufactured. The expectation is that $f_\mathrm{diff} = \sqrt{m_\mathrm{e} / m_*}$\cite{green-1990}, and typically the effective mass of an electron in a silicon lattice is $m_* \sim 0.28\,m_\mathrm{e}$, giving $f_\mathrm{diff}\sim 1.9$. Comparing \textsc{poisson\_ccd} simulations with lab measurements of CCD devices yields values of the diffusion multiplier between 1.4 and 2.3\cite{lamarr+2022}. In simulating charge diffusion for this work, we adopt a value of $f_\mathrm{diff} = 1.7$.

While the energy of a photon determines the depth in the silicon at which it will be absorbed, the dominant factor in determining the final cloud radius is only the depth at which it is absorbed (\textit{i.e.} photons of different energies absorbed at the same depth produce charge clouds of the same size at readout). We find that the predicted cloud sizes as a function of energy and interaction depth to be in good agreement between the full \textsc{poisson\_ccd}, with a consistent value of the diffusion multiplier between the simulation and the analytic model. 

\begin{figure}
    \centering
    \subfigure[] {
    \label{fig:sims:1}
    \includegraphics[width=0.23\linewidth]{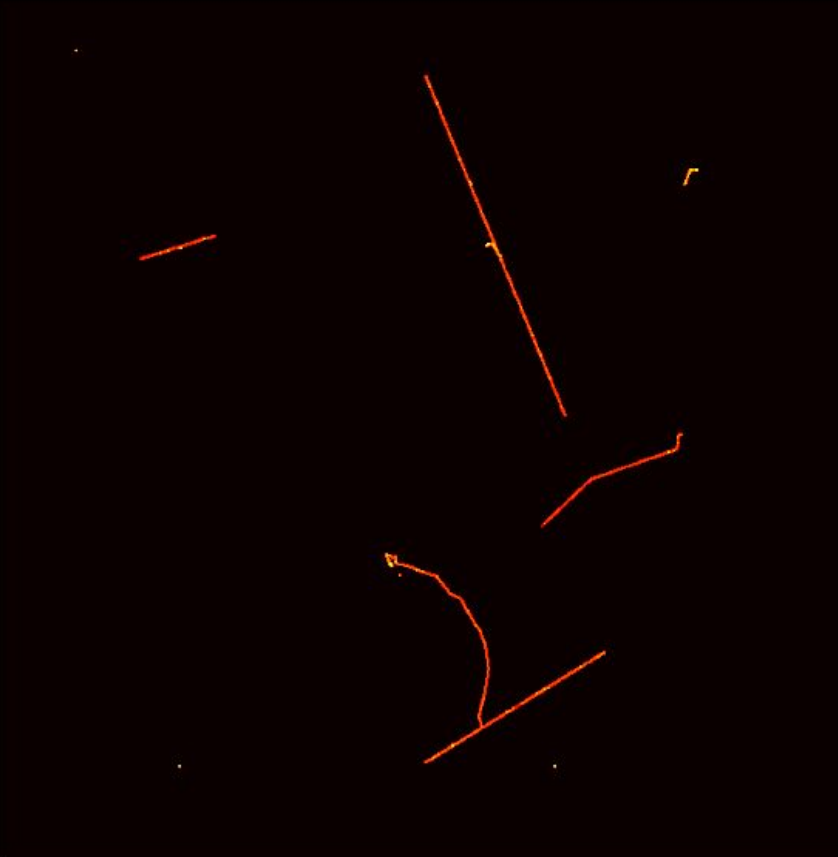}
    }
    \subfigure[] {
    \label{fig:sims:2}
    \includegraphics[width=0.23\linewidth]{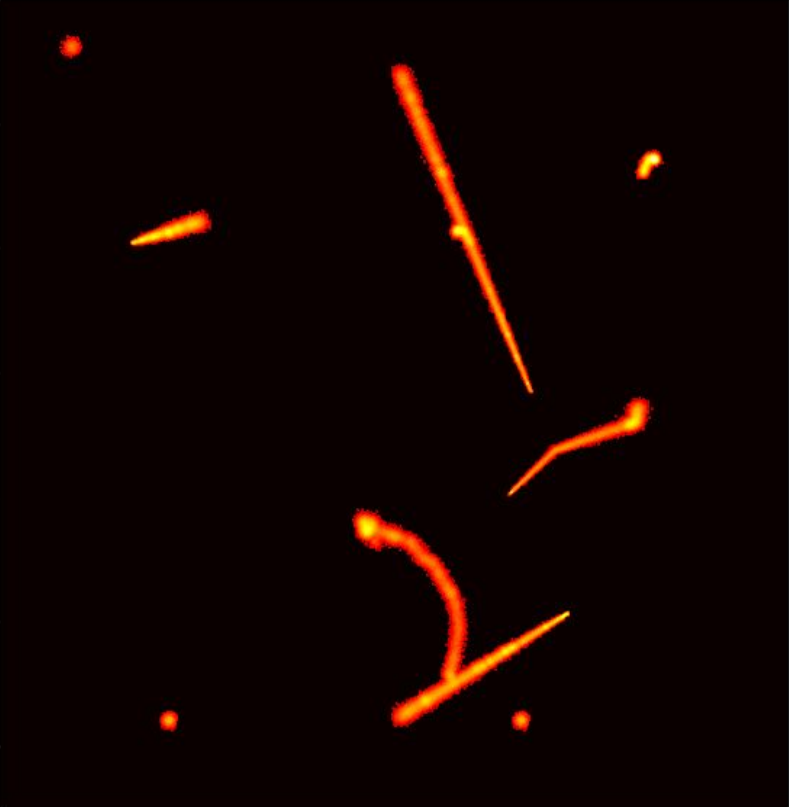}
    }
    \subfigure[] {
    \label{fig:sims:3}
    \includegraphics[width=0.23\linewidth]{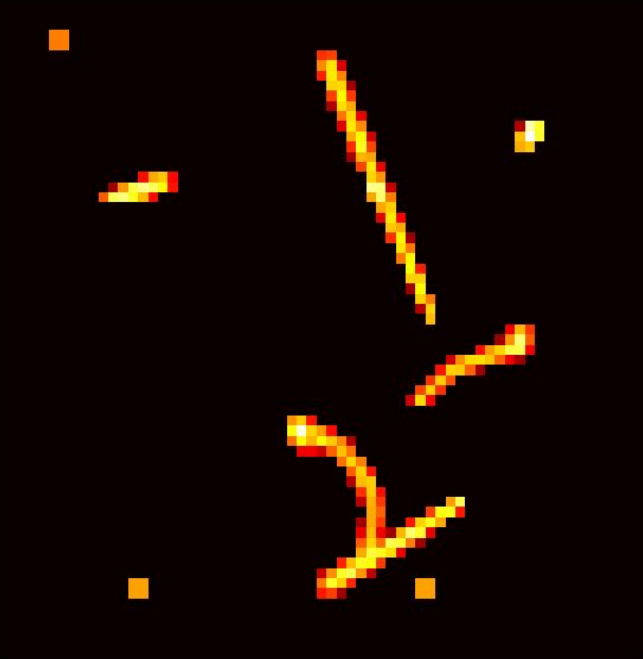}
    }
    \subfigure[] {
    \label{fig:sims:4}
    \includegraphics[width=0.23\linewidth]{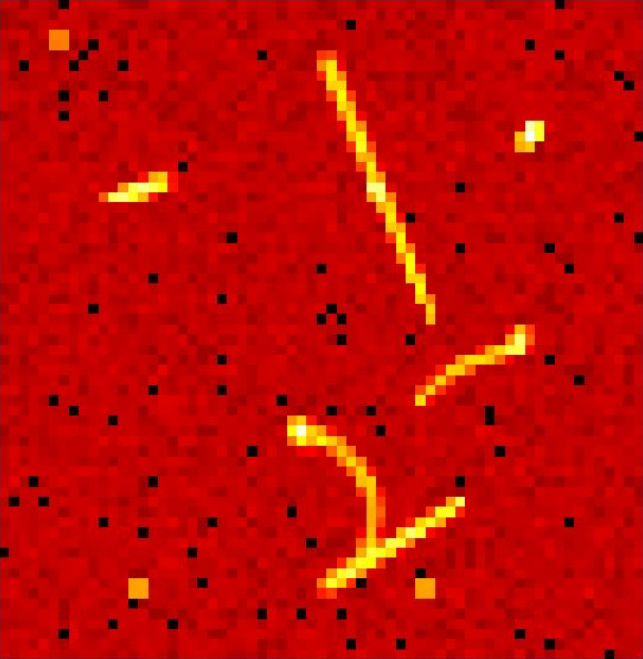}
    }
    \caption{Example of a simulated detector frame through the stages of the pipeline. \subref{fig:sims:1} Initial energy depositions as a function of $(x,y)$ position within the detector from both astrophysical X-ray photons, and cosmic particles and their secondaries, simulated using \textsc{geant4}. \subref{fig:sims:2} Positions of the electrons liberated within the detector as energy from photons and charged particles is deposited, following diffusion as the electrons drift through the depth of the detector to the readout gates. \subref{fig:sims:3} The number of electrons within each of the detector pixels. \subref{fig:sims:4} The final simulated frame, including readout noise.}
    \label{fig:sims}
\end{figure}

Following diffusion of the electron clouds as they drift to the depth of the readout gates, the electrons are sorted into pixels. A square grid is drawn over the detector representing the pixel boundaries and it is assumed that any electron within the boundaries of a pixel is recorded within that pixel. Finally, readout noise is added to the frame, assigning a value to each pixel drawn from a Gaussian distribution, with mean corresponding to the number of electrons in the pixel and standard deviation corresponding to the RMS noise level (using a representative value three electrons RMS noise for the \textit{Athena WFI} and \textit{AXIS} CCDs). An example of a simulated frame is shown in Figure~\ref{fig:sims}.

\section{AI filtering of cosmic ray-induced background signals}
The limitation of traditional background filtering methods, based upon the total energy of an event being below a defined threshold, and the pattern or grade of the illuminated pixels being valid (\textit{i.e.} single or double-pixel events, or triple- and quad-pixel events in valid patterns), is that each event is considered in isolation, and the context within which each event appears is not taken into account. A significant number of background events appear to be valid in isolation, where small amounts of energy are deposited into isolated pixels. Indeed, many valid background are events \textit{are} X-ray photons themselves, produced as the charged particle interacts with the material of the spacecraft. For example, it may be possible to identify an otherwise-valid background event if it appears in close proximity to an extended particle track, if the particles creating both the track and valid event originate within a single shower of secondaries produced from one cosmic ray interaction. The proposed \textit{self-anti-coincidence} (SAC) method addresses such events by defining an exclusion radius around any extended track that appears within a frame. However, optimizing this exclusion radius is non-trivial and enhanced background filtering comes at the expense of rejecting valid astrophysical X-ray photons\cite{miller_jatis}.

It may also be possible to identify and filter valid background events that do not appear alongside extended particle tracks or high-energy particle events, but rather appear alongside other valid events. One of the primary advantages of operating next-generation X-ray detectors at high frame rates is that the number of events (cosmic ray-induced or astrophysical) expected per readout frame is small. It is planned to read out one full frame of the \textit{Athena} WFI detector every 5\ms\ (or indeed every 2\ms\ with more recent developments of the \textit{NewAthena} concept); thus, approximately one cosmic ray event, and less than one astrophysical photon event for low surface brightness sources, is expected per quadrant of the detector in each frame\cite{grant+2020}. Therefore, if many `valid' background events appear in close proximity to one another, it will be possible to identify them as likely having originated from the same shower of secondaries produced by a charged particle.

We investigate whether enhanced background filtering can be performed by applying machine learning algorithms to the frame-by-frame data that are read from the detector. The algorithms we are investigating are based upon convolutional neural network (CNN) architectures that are a cornerstone of image recognition applications in AI. CNN algorithms are composed of a number of layers of convolutional filters that are fit to the specific image features relevant to the identification problem. Between these layers, a series of pooling operations are performed to down-sample the image to lower resolutions. This results in successive convolutional filters operating on features of increasing size within the image, to first identify the small-scale features, then put them into the context of the larger-scale features. The output of these convolution and pooling layers can then be passed to network layers that either classify the entire image (\textit{i.e.} a fully-connected or dense layer), or to a series of layers that again up-sample the classifications to identify specific regions or objects within the image, or to classify the contents of the image on a pixel-by-pixel basis, as is performed in the popular \textsc{unet} architecture\cite{unet}.

We train the neural network algorithms using a library of simulated frames that would be read out of the detector. Each frame contains signals originating from cosmic ray particles (drawn from the library of \textsc{geant4} simulations) and from astrophysical X-rays that are drawn from a uniform spatial distribution, with a flat spectrum in the 0.3-10\keV\ band. A separate library of simulated frames (which are not included in the training set) are used to test the performance of the algorithms and to measure the degree to which they are able to identify background \textit{vs.} astrophysical events that they have not seen before.

By training machine learning algorithms on simulated X-ray and cosmic ray events, they can learn the relevant physical correlations between particle tracks and `valid' background events that enable enhanced identification and filtering. Such an algorithm offers key advantages over the self-anti-coincidence method in that there is not a need to tune an \textit{ad hoc} exclusion radius (the correlation scales can be learned from the training data), and, in principle, `valid' background events can be identified by their correlations with one another, rather than needing to be associated with a particle track in the same frame. Classification or `prediction' values produced by CNN algorithms can be interpreted as probabilities, \textit{i.e.} the probability that a given signal is induced by the background. These probabilities for each event can be added to the event list, and the end user can determine the threshold value at which an event is excluded from the final science analysis.

We treat each frame read from the detector after a single integration time as a single image that is input to the CNN algorithms (or, in the case of the \textit{Athena WFI}, which employs a rolling shutter theme, reading continually, line-by-line, we consider one sequence of lines comprising the full frame from one of the four $512\times 512$ pixel quadrants as a single image frame). One of the key advantages of next-generations detectors will be the high frame rate and short integration time. This means that the number of events per frame will be small, with approximately one cosmic ray event per quadrant of the WFI detector per frame\cite{grant+2020}, and fewer than one X-ray photon per frame from the faint astrophysical sources for which effective background filtering will be small. This places important constraints on the classification of events, since it is unlikely that multiple X-rays will appear int he same frame, enabling showers of `valid' events from secondaries produced by the same cosmic ray primary to be identified more easily based upon the spatial correlations between these events. If the frame rate were lower, spatial correlations between genuine X-ray events would make it more difficult to identify such secondaries.

For the purposes of training the neural network from event simulations, we consider all signals induced by a cosmic ray and secondaries that are produced as it interacts with the spacecraft and detector to fall into the cosmic ray classification. These secondaries will include X-ray photons that are produced in interactions with material on the spacecraft that to all intents and purposes are X-ray photons and, in isolation, would be classified as such. By classifying such photons as cosmic ray events, the algorithm may learn how to identify these background X-ray signals via their correlation with other particle-induced events that appear in the same frame.

\subsection{First generation algorithm: classification of detector frames}
The first generation machine learning algorithm that was tested performed simple classification of $64\times 64$ pixel subframes to demonstrate the feasibility of applying AI to identify and filter the instrumental background\cite{spie_2020}. This algorithm consists of two layers of convolutional filters, followed by pooling, then classification. For each input frame, the output consists of three numbers, corresponding to the probability that the frame contains (1) only genuine astrophysical X-rays, (2) only cosmic ray-induced signals, or (3) both X-ray and cosmic ray signals. 

This proof-of-concept algorithm was able to correctly identify 99 per cent of frames that contain a cosmic ray event, and was able to identify 40 per cent of the cosmic ray-containing frames that are missed by traditional filtering methods. Where a frame contains both X-ray and cosmic ray events, 97 per cent of test frames were correctly identified as containing both, showing that CNN algorithms can, in principle, separate cosmic ray and X-ray signals from the same frame. Moreover, this frame classification algorithm correctly identified `valid' background events, including X-ray photons and low energy electron or positron events, when they appear alongside other events within the same frame, demonstrating that the neural network is able to identify the context and correlations between valid events that originate from the same cosmic ray primary.

\subsection{Second generation algorithm: sliding-window identification of regions of interest}
The second generation algorithm\cite{spie_2022} was developed to extend AI identification of background events from $64\times 64$ pixel subframes to full $512\times 512$ quadrants of the \textit{Athena WFI detector}. When working with such a large detector area, it is not sufficient to identify and reject an entire frame containing cosmic ray signals, since the probability that a frame will contain at least one cosmic ray signal is high\cite{miller_jatis,grant+2020}. It is necessary to separate the regions of the detectors containing the cosmic ray and astrophysical X-ray signals.

This algorithm operates in two stages. Initially, $64\times 64$ pixel regions are extracted from the full $512\times 512$ frame using a sliding window, with some defined step size. The non-empty windows are then passed to a CNN classification algorithm to identify (with a probability value between 0 and 1) these regions of interest as containing either only astrophysical X-rays, (2) only cosmic ray-induced signals, or (3) both X-ray and cosmic ray signals. The sliding window scheme will naturally produce windows that overlap. In this case, the prevailing classification for any given region of the detector is taken from the window for which the prediction value in any of the categories is the highest. For any given event, such a scheme has the effect of incorporating the context from any surrounding signals to yield the extremal prediction value for that event (\textit{i.e.} in deciding whether an event is part of the cosmic ray-induced background, the highest probability is computed for when it is taken in the context of \textit{any} of the surrounding events).

In the most aggressive mode of background filtering, events are excluded from any $64\times 64$ pixel region of interest that is identified as containing either only cosmic ray signals or a mixture of cosmic ray and X-ray signals. In this case, we find that the AI algorithm is able to reduce the unrejected instrumental background by 30 per cent relative to traditional filtering methods. Such aggressive filtering, rejecting events from any $64\times 64$ pixel region identified as containing a background event, inevitably leads to a reduction in the active area of the detector for that frame, so will come at the expense of lost X-ray signals. At the high frame rates proposed for next-generation X-ray detectors (5\ms\ frame integration times), we expect that only around 5 per cent of the genuine X-ray signals from faint astrophysical sources (for example the outskirts of nearby galaxy clusters) will be lost, since the probability of an X-ray photon landing in the same region of the detector during such a short time window is small. 96 per cent of the `valid' background events that were successfully identified as resulting from cosmic rays were found in frames with at least one other event, supporting the conclusion that the primary gain of the AI algorithms over traditional filtering algorithms is the ability to consider events in the context of the other events with which they appear, and by learning the spatial correlations that appear between events resulting from the same primary cosmic ray particle.

A more conservative mode of filtering is available, passing the regions of interest identified as containing both cosmic ray and X-ray signals to a secondary neural network algorithm that identifies the locations of the cosmic ray and X-ray events within the frame. For this purpose, we employed a \textsc{unet} algorithm, which classifies the image frame pixel-by-pixel, assigning a probability that each pixel belongs to either an X-ray or cosmic ray-induced event. We then reconstruct the valid events from the $3\times 3$ pixel islands surrounding the local maxima using the traditional method, and assign to each event the maximum cosmic ray prediction value (\textit{i.e.} the probability that the event is due to a cosmic ray) from the \textsc{unet} from the nine pixels within the island. Due to limitations with the current \textsc{unet} architecture, this conservative mode of filtering only provided a 6 per cent reduction in the unrejected background, compared to traditional filtering methods.

\subsection{Third generation algorithm: classification of events}
\begin{figure}
    \centering
    \includegraphics[width=0.75\linewidth]{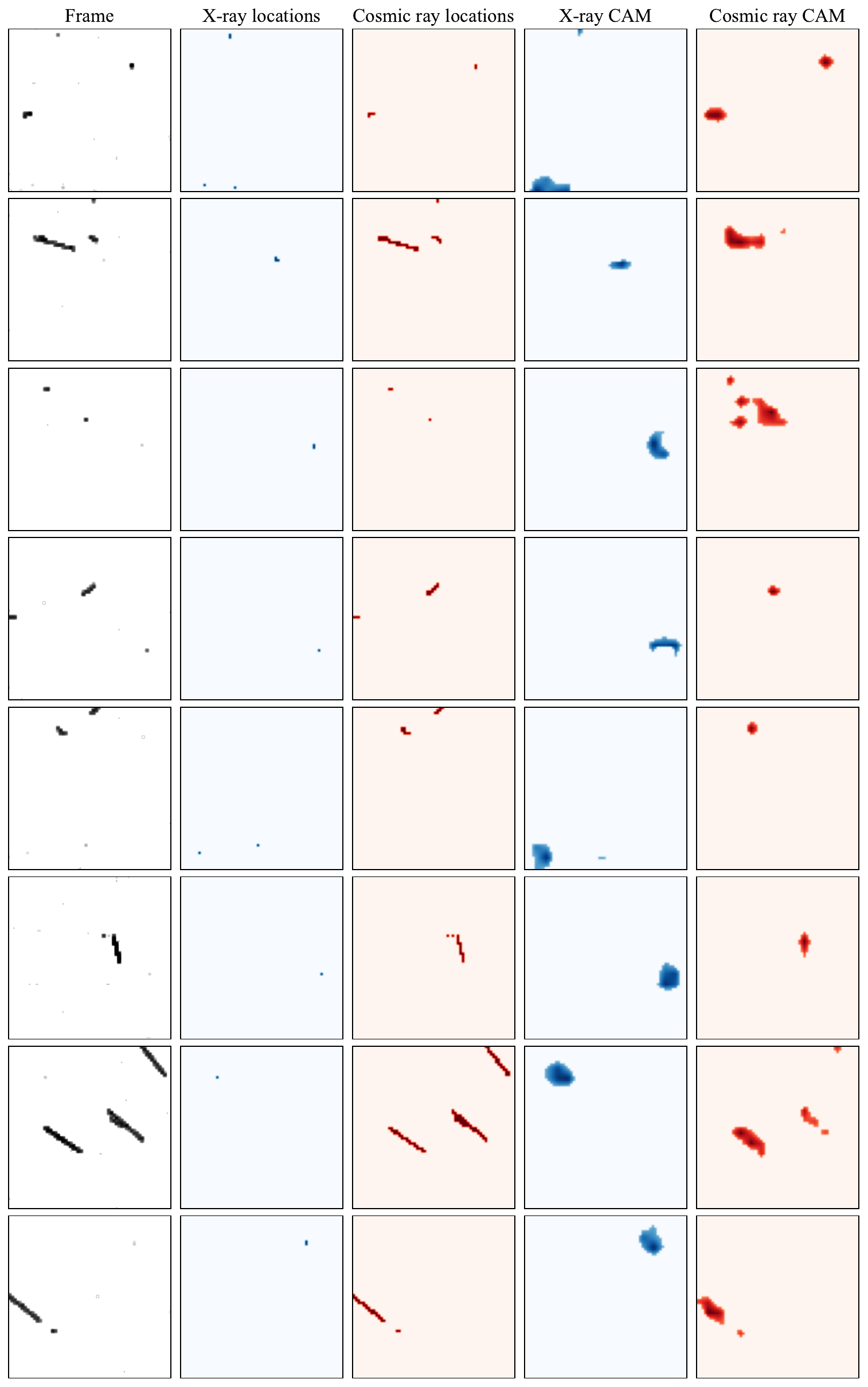}
    \caption{Operation of the third generation AI algorithm to identify cosmic ray-induced background events in individual detector frames. The left panel shows the simulated frame, where the shading represents in the recorded pixel values. The second and third panels show the true locations of X-ray and cosmic ray-induced events in these frames. The fourth and fifth panels show the class activation map (CAM) values from the convolutional neural network (CNN), highlighting the locations in the frame in which genuine X-ray and cosmic ray-induced background events have been identified.}
    \label{fig:cams}
\end{figure}

The latest generation of the algorithm was designed to specifically address the problem of separating genuine, astrophysical X-ray events from cosmic ray-induced background events within the same frame. This algorithm is based upon the CNN frame classification algorithm, which uses a series of convolutional filters to detect features in the image frame, producing a set of \textit{feature maps} where each filter is activated within the image. The output of these filter layers is then passed to fully connected or dense layers that classify the entire frame as containing either only astrophysical X-ray signals, only cosmic ray signals, or a mixture of both. Once a frame has been classified, the third generation algorithm computes \textit{class activation maps} (CAMs)\cite{gradcam,gradcam++}, which highlight the regions of the image containing the features that lead the frame to be classified in a certain way (Fig.~\ref{fig:cams}). A CAM is produced for each of the three classes, showing the specific features within the image that would pull the classification in those directions.

The raw CAM values produced by the neural network algorithm are non-trivial to interpret directly, so a \textit{random forest classifier}, which is trained on the same data set, is used classify each pixel as belonging to (1) an astrophysical X-ray event, or (2) a cosmic-ray induced background event. Each event is defined in a $3\times 3$ island around the local maximum pixel value as per the traditional event reconstruction methods, and the pixel-wise CAM values are aggregated by taking the minimum, maximum and median values within a $5\times 5$ window centred on the event (adding the additional context from the pixels surrounding the original $3\times 3$ island). These aggregated CAM values are input into the random forest classifier along with the total energy of the event (reconstructed by summing the pixel values that are above the split threshold) and the summary classification value for the entire frame from the CNN.

\begin{figure}
    \centering
    \subfigure[] {
    \label{fig:ai_performance:filt}
    \includegraphics[width=0.48\linewidth]{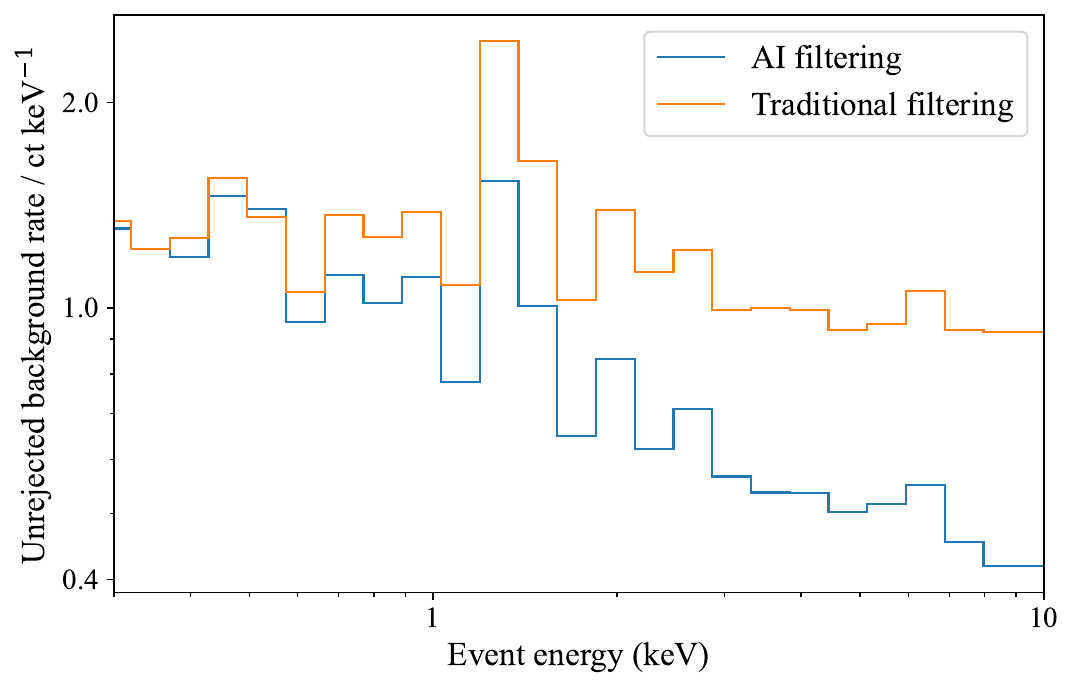}
    }
    \subfigure[] {
    \label{fig:ai_performance:lost}
    \includegraphics[width=0.48\linewidth]{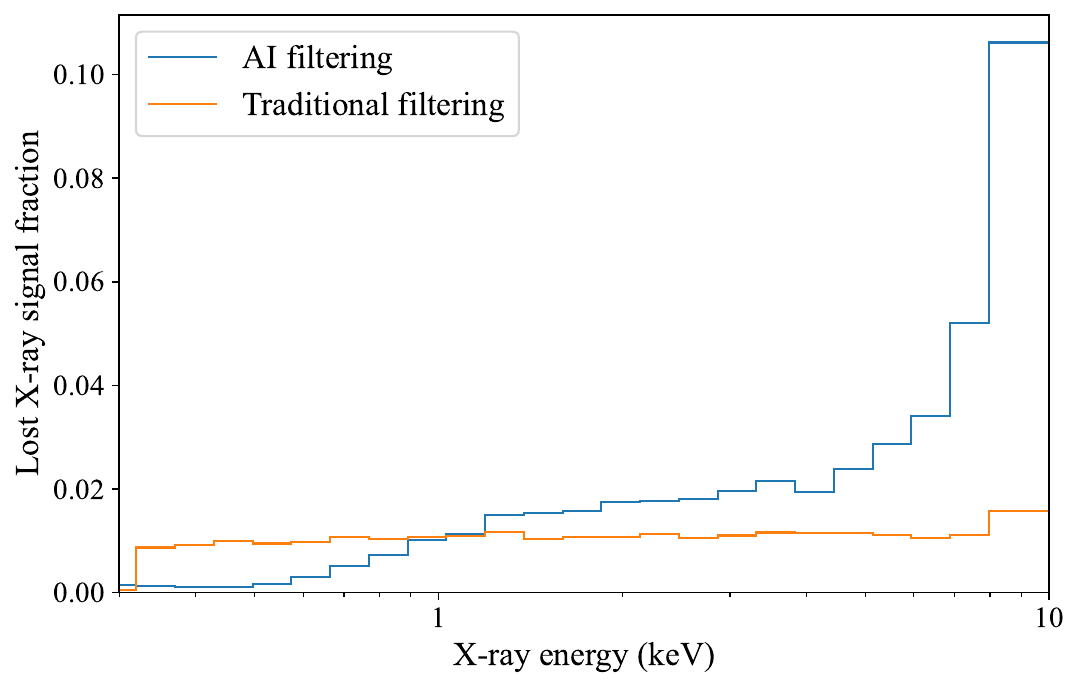}
    }
    \caption{Performance of the third generation AI algorithm, showing \subref{fig:ai_performance:filt} the remaining cosmic ray-induced instrumental background spectrum following filtering by the AI algorithm, compared with that following traditional background filtering methods (selecting single, double, triple and quadruple pixel events with valid patterns and energies, with the additional filtering provided by the $5\times 5$ editing or `very faint mode'), and \subref{fig:ai_performance:lost} the fraction of genuine, astrophysical X-ray events, as a function of energy, that are incorrectly removed by filtering with the AI algorithm and with traditional filtering methods. Simulations are performed for an \textit{Athena WFI}-like detector, with instrumental background events derived from \textsc{geant4} simulations of interactions between cosmic ray protons and the spacecraft and detector.}
    \label{fig:ai_performance}
\end{figure}

We find that the third generation AI algorithm, operating on $64\times 64$ pixel subframes, is able to reduce the unrejected background signal by 41.5 per cent across the 0.3-10\keV\ energy band (Fig.~\ref{fig:ai_performance}a), compared to traditional background filtering methods, and by 45.1 per cent over the 4-8\keV\ band, which is of particular interest for a wide range of science cases that can be conducted using measurements of the iron K line. We find that this enhanced filtering comes at the expense of a slight increase in the fraction of true X-ray events that are incorrectly rejected, although this remains within a a tolerable level, with only 1.8 per cent of true X-ray events lost in the 0.3-10\keV\ band. 

We compare the performance to the performance of the AI algorithm to the the modified \textit{ASCA} grading criteria employed in the data reduction pipelines the \textit{Suzaku XIS}, and more recently the \textit{XRISM Xtend} CCDs, in which valid events are considered to be single, double, triple and quadruple pixel events with valid patterns. In the traditional filtering method, we additionally include  $5\times 5$ editing mode (also known as the `very faint mode'), designed to provide enhanced background filtering of valid $3\times 3$ events that appear close to charged particle events. In this mode, when a triple or quadruple pixel event is detected in a $3\times 3$ island, event is rejected if any additional signal is detected in the surrounding 16 pixels, forming a $5\times 5$ island around the event.

\begin{figure}
    \centering
    \includegraphics[width=0.5\linewidth]{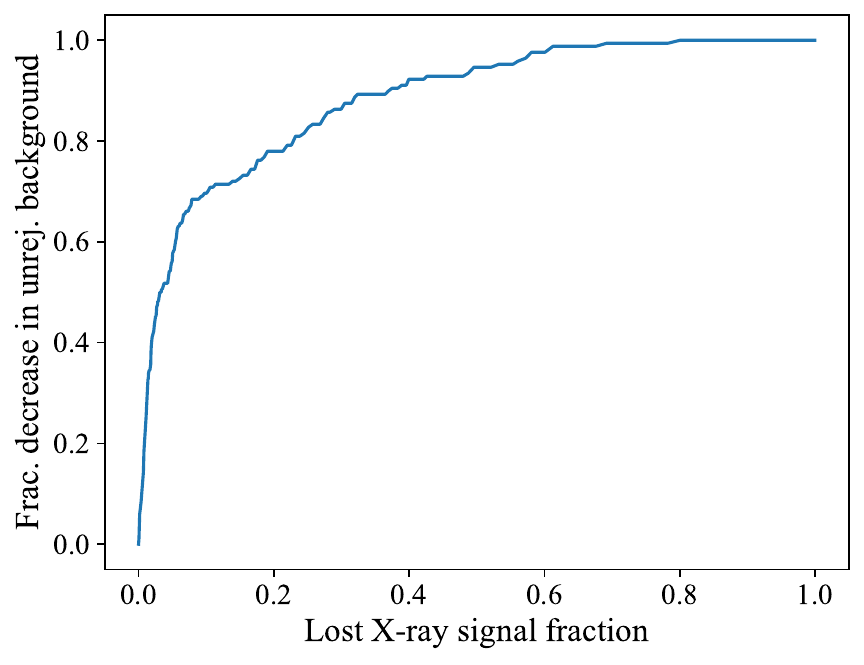}
    \caption{Receiver operating characteristic (ROC) curve for the random forest classifier component of the third generation AI filtering algorithm, showing the relationship between the fractional decrease in unrejected background \textit{vs.} the lost genuine X-ray signal as the background rejection threshold is tuned. The third-generation algorithm is able to reduce the unrejected background signal by 41.5 per cent across the 0.3-10\keV\ energy band, losing just 1.5 per cent of the valid X-ray signal.}
    \label{fig:roc}
\end{figure}

It is possible to tune the output of the random forest classifier, adjusting the threshold value at which an event is considered to be a cosmic ray or X-ray. While the above performance are for representative figures, it is possible to adjust the threshold to reject a greater proportion of the unrejected background, though this comes at the expense of greater lost X-ray signal. Fig.~\ref{fig:roc} shows the fractional decrease in the unrejected background \textit{vs.} the fraction of genuine X-ray events that are incorrectly rejected that can be achieved by tuning the classifier for the third generation algorithm.

Further details about the third generation algorithm and its performance are presented in Poliszczuk et al. 2024\cite{poliszczuk+2024}.

\section{Enhanced reconstruction of photon events}
Once a photon (or cosmic ray particle) is absorbed in the silicon medium of the detector, a cloud of electrons is liberated which must then drift, through the electric field, to the readout gates, before the signal is recorded. Random diffusion of the electrons causes this cloud to spread as it drifts towards the gates, and depending upon the size of the cloud relative to the size of the pixels (\textit{i.e.} the separation between the readout gates), and the location of the initial interaction point with respect to the pixel boundaries, a significant portion of this charge cloud can be spread across multiple pixels, creating \textit{split events}. It is therefore necessary to \textit{reconstruct} each event and calculate the initial photon energy from the total number of electrons that ends up in each pixel.

In traditional event reconstruction methods\cite{xis}, each event is defined as a $3\times 3$ island of pixels around each local maximum pixel value (above a threshold known as the \textit{event threshold}). The total event energy is calculated by summing the pixel values within the $3\times 3$ island that exceed a second threshold, known as the \textit{split threshold}, usually defined to be above the $3\sigma$ limit for the distribution of noise in the empty pixels. For comparison purposes, we here adopt a split threshold of 80\eV\ within a pixel, corresponding to 22 electrons. A crucial limitation of this method is that a significant fraction of the event energy can be carried by electrons into pixels that do not rise above the split threshold, especially for the lowest energy photon events. This fraction of the event energy is lost, resulting in a bias in the recorded energy of these photons, and a broadening of the spectral response function.

We investigate an alternative event reconstruction method, based upon fitting a model to the recorded pixel values to recover the signal that is lost into the pixels that do not reach the requisite threshold\cite{lamarr+2022,jones+2024}. The model is based upon a two-dimensional Gaussian representing the spatial distribution of the electrons after they have drifted and diffused from the interaction point to the side of the device bearing the readout gates (we do not include the final part of the drift where electrons are pulled towards the readout gate in the centre of each pixel, but rather assume that any electron within the boundaries of a pixel a small distance above this plane will be counted by the gate within the pixel). The spatial density of electrons at positions $(x,y)$ across the readout plane follows:

\begin{equation}
\label{equ:gauss_model}
    n_\mathrm{e}(x,y)\,dx\,dy = \frac{N_\mathrm{e}}{2\pi\sigma} e^{-\frac{1}{2\sigma^2}\left[(x-x_0)^2 + (y-y_0)^2\right]}\,dx\,dy
\end{equation}
The total number of electrons within the cloud, $N_\mathrm{e}$ is drawn from a Fano distribution dependent upon the energy absorbed from the photon (or particle), with an average of one electron liberated per 3.65\eV\ of energy deposited in silicon.

\begin{figure}
    \centering
    \includegraphics[width=0.5\linewidth]{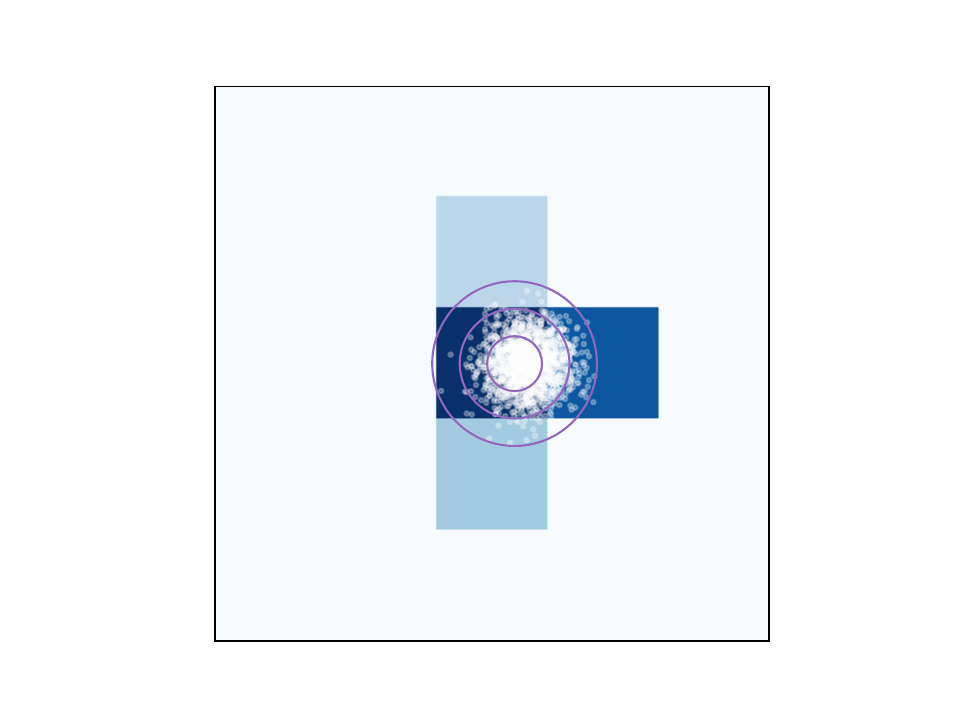}
    \caption{Illustration of the Gaussian fitting event reconstruction method for a 6.4\keV\ photon in an \textit{AXIS}-like detector, with 24\um\ width by 100\um\ depth pixels. White dots show the location of individual electrons after they have diffused to the plane within the detector holding the readout gates (using the analytic diffusion model, showing positions before the electric field draws the electrons into the gate itself). Shading denotes the value recorded in each pixel (\textit{i.e.} the number of electrons falling within the pixel boundary, representing the actual data available from the detector). Purple contours show the best-fitting two-dimensional Gaussian model for the electron cloud, fit via least-squares minimisation to the values recorded in the pixels (denoted by shading). Lines represent the 1, 2 and $3\sigma$ radii.}
    \label{fig:gauss_fit}
\end{figure}

The model for the number of electrons expected to be recorded in each pixel is computed by integrating this distribution over the area of each pixel:
\begin{align*}
    N_{ij} &= \int_{x_i}^{x_{i+1}}\int_{y_j}^{y_{j+1}}\frac{N_\mathrm{e}}{2\pi\sigma}e^{-\frac{(x-x_0)^2}{2\sigma^2}}e^{-\frac{(y-y_0)^2}{2\sigma^2}}\,dx\,dy\\
    \label{equ:int_model}
    & = N_\mathrm{e} \left[\mathrm{erf}\left(\frac{x_{i+1} - x_0}{\sqrt{2}\sigma}\right) - \mathrm{erf}\left(\frac{x_{i} - x_0}{\sqrt{2}\sigma}\right)\right]\left[\mathrm{erf}\left(\frac{y_{j+1} - y_0}{\sqrt{2}\sigma}\right) - \mathrm{erf}\left(\frac{y_{j} - y_0}{\sqrt{2}\sigma}\right)\right] \numberthis
\end{align*}
Where the free parameters of the model are $(x_0, y_0)$, the central position of the cloud, corresponding to the location at which the photon was absorbed, $N_\mathrm{e}$, the total number of electrons in the cloud, and $\sigma$, the radius of the cloud defined by the $1\sigma$ width of the Gaussian. $\mathrm{erf}(x)$ is the error function, defined as the integral of the Gaussian distribution from $-\infty$ to $x$. The energy of the event is calculated from the number of electrons: $E(\mathrm{eV}) = 3.65/N_\mathrm{e}$. The model is fit to the pixel values recorded in the $5\times 5$ pixel grid surrounding the local maximum pixel of the event using a simple least-squares minimisation routine, yielding the best-fitting parameter values for each event. The Gaussian fitting event reconstruction scheme is illustarated in Fig.~\ref{fig:gauss_fit}.

While it is possible to model each pixel value directly using the Gaussian distribution in Equation~\ref{equ:gauss_model} in the case that the cloud radii are much larger than the pixels (as was performed in previous work\cite{lamarr+2022}), we find that it is necessary to model the integrated charge in each pixel using Equation~\ref{equ:int_model} when the cloud radius is smaller than the pixel size, as will be the typical case for CCD or DEPFET devices. The former model fails to describe split events when the event centroid lies close to the boundary between pixels and the charge is split more-or-less equally between the pixels. The value of the Gaussian in the centre of the pixels in this case will have dropped significantly from the peak value that falls between the pixels. Modelling the integrated charge in each pixel accounts for all of the electrons that fall within each pixel wherever the centroid lies.

We test this event reconstruction scheme on a set of simulated X-ray photon events for a range of detector characteristics. The most relevant detector parameters determining the diffusion of signals between pixels are the size of the pixels, and the depth of the pixels between the entrance window and the readout gates. The method was tested for detector models representing the \textit{Athena WFI} (130\um\ width $\times$ 450\um\ depth pixels) and the \textit{AXIS} focal plane CCDs ($24\times 100$\um\ pixels). Our calculations show that the ratio between the charge cloud radius at the readout plane and the width of the pixels is approximately the same between the two detector designs, therefore the performance of the event fitting method is largely consistent between them.

\begin{figure}
    \centering
    \includegraphics[width=1\linewidth]{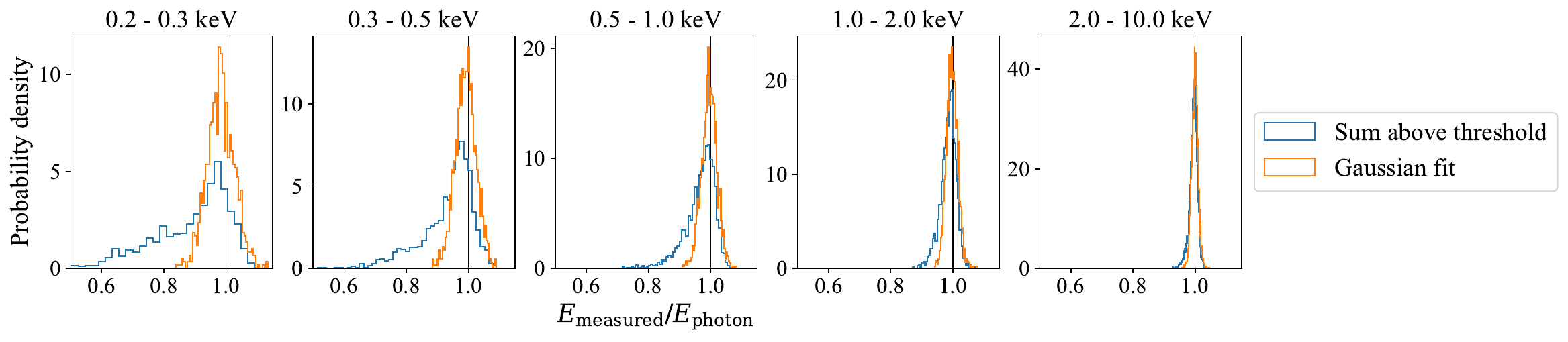}
    \caption{Spectral line responses for the sample of simulated X-ray events in an \textit{AXIS}-like CCD, showing the distribution of the radio of the measured to input photon energy for all events in the sample, divided into energy channels. The energy of each event is measured using both the traditional method, summing pixels above threshold, and the Gaussian reconstruction method, fitting a model of the charge cloud to the measured pixel values. For events below 1\keV, a significant fraction of the energy is carried into surrounding pixels that do not rise above the split threshold, thus are discarded. This leads to both a shift in the centroid of the event energy, biasing the measured energy of low energy photons, and produces an extended low energy tail on the line response function, increasing the uncertainty on the energy of these events. The Gaussian reconstruction method is able to recover this energy that is lost from the events, producing unbiased photon energy measurements with significantly smaller uncertainty. Charge diffusion and event characteristics are similar in an \textit{Athena WFI}-like device.}
    \label{fig:rmf}
\end{figure}

\begin{figure}
    \centering
    \subfigure[] {
    \label{fig:event_fit_energy:energy}
    \includegraphics[width=0.48\linewidth]{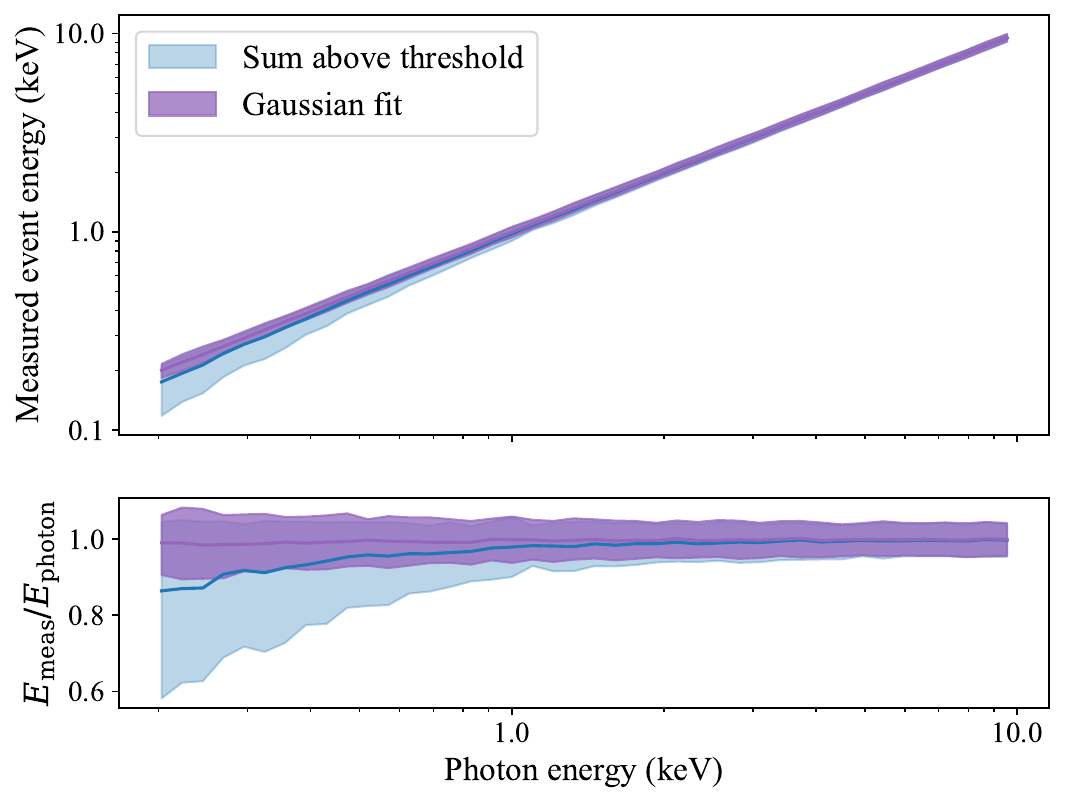}
    }
    \subfigure[] {
    \label{fig:event_fit_energy:fwhm}
    \includegraphics[width=0.48\linewidth]{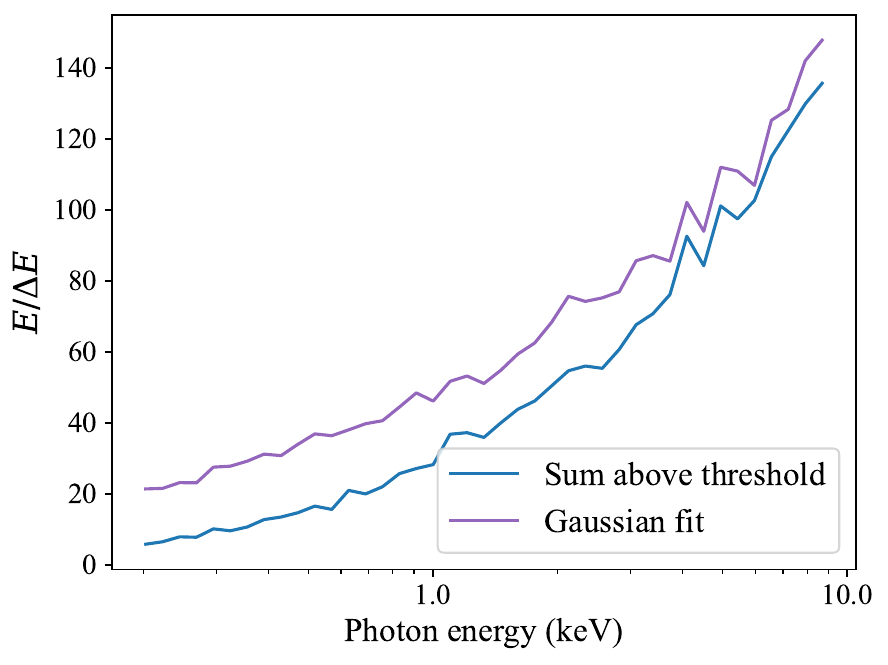}
    }
    \caption{Performance of the Gaussian event reconstruction method, for simulated X-ray photon events in an \textit{AXIS}-like CCD (charge diffusion and event characteristics are similar in an \textit{Athena WFI}-like device). \subref{fig:event_fit_energy:energy} The distribution of measured event energies as a function of input photon energy, using the traditional method, summing pixels above threshold, and the Gaussian reconstruction method, fitting a model of the charge cloud to the measured pixel values. Contours represent 90 per cent of the measured values. \subref{fig:event_fit_energy:fwhm} The energy resolution, $R = E / \Delta E$, where $\Delta E$ is defined as the full width at half maximum (FWHM) of the line response function, as a function of photon energy using the traditional and Gaussian reconstruction methods. Simulations do not include readout noise, thus represent only the effects of charge diffusion between pixels on the energy resolution.}
    \label{fig:event_fit_energy}
\end{figure}

Simulations of the charge diffusion between pixels show that the energy reconstruction of photons below 1\keV\ can be severely impacted by the loss of signal into neighboring pixels. For each input photon energy, a distribution of event energies is measured (often described as the photon redistribution matrix or RMF). Employing the traditional event reconstruction method of summing the signals recorded in pixels above the split threshold leads not only to broadening of the distribution, increasing uncertainty in the photon energy at low energies, but also a shift in the centroid of the distribution, representing the absolute energy scale calibration (Fig.~\ref{fig:rmf}), as energy from these events is lost into neighboring pixels that for the lowest energy photons do not rise above the threshold. The Gaussian reconstruction method, fitting the measured signal in each pixel with a model of the diffused charge cloud is able to account for the energy lost into these pixels, and by fitting the peak of the signal in the pixels that exceed the threshold, is able to correctly recover the correct centroid energy for the photon and reduce the uncertainty in its measurement. Fig.~\ref{fig:event_fit_energy:energy} summarises how the distribution of measured event energies varies as a function of input photon energies and Fig.~\ref{fig:event_fit_energy:fwhm} shows the uncertainty on the event energy (expressed as the FWHM of the spectral response).

These simulations do not currently include noise introduced in the readout of the pixel values, and as such represent only the effects of charge diffusion between pixels on the event reconstruction and energy resolution of the detector. The addition of read noise will increase the uncertainty on the reconstructed event energies. In principle, read noise can be readily incorporated into the Gaussian reconstruction scheme via the likelihood function that compares the model with the recorded pixel values,(a $\chi^2$ statistic in the case of Gaussian noise), then replacing the least squares minimisation with minimisation of that likelihood function.

\subsection{Distinguishing X-ray and cosmic ray events using Gaussian reconstruction}
In addition to reconstructing the energy of each event, we might also consider whether additional information is available in the characteristics of each event, derived from fitting a model to the resulting charge cloud in the detector, which can help distinguish genuine astrophysical X-ray events from cosmic ray-induced background events.

\begin{figure}
    \centering
    \subfigure[] {
    \label{fig:en_radius:xray}
    \includegraphics[width=0.48\linewidth]{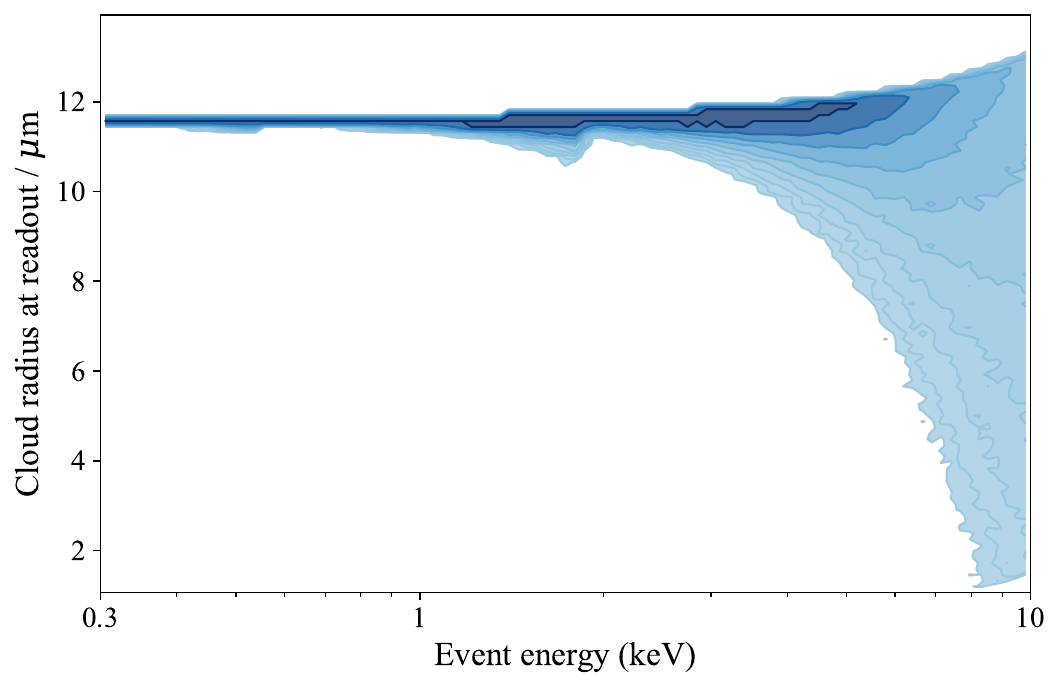}
    }
    \subfigure[] {
    \label{fig:en_radius:gcr}
    \includegraphics[width=0.48\linewidth]{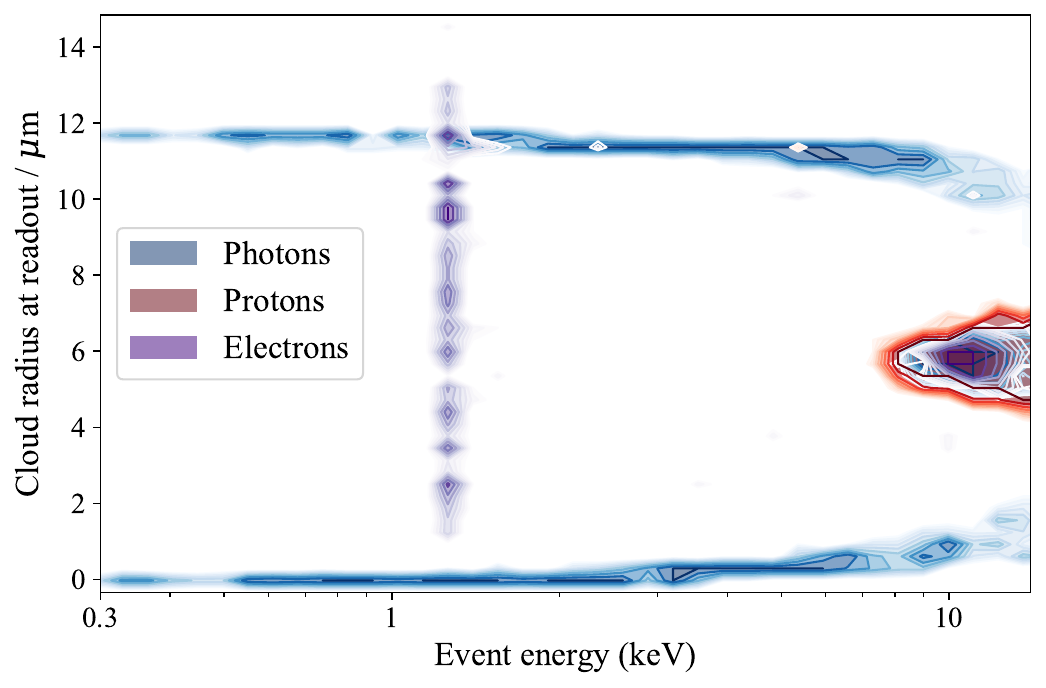}
    }
    \caption{Simulated distribution of the charge cloud radius for events within a silicon \textit{Athena WFI}-like detector, after the electrons have drifted and diffused from the location of energy deposition towards the readout gates. Cloud radii are calculated using the analytic diffusion model, based upon the depth at which the energy is deposited, for \subref{fig:en_radius:xray} astrophysical X-ray photon events, entering the detector via the entrace window, and \subref{fig:en_radius:gcr} `valid' cosmic ray-induced background events that would not be rejected by traditional filtering methods. Valid events are defined as those in which the total energy deposited is below 10\keV\ and those in which the energy deposition spans over no more than four pixels, so as to produce valid event patterns. Shading represents the type of secondary particle from the cosmic ray event that deposits energy within the detector. It should be noted that the where energy deposition is extended through the depth of the detector, the weighted mean depth is computed, thus the island of proton-dominated events appearing at intermediate cloud radii at energies above approximately 10\keV\ in fact represent extended clouds that result from depositions across a range of depths.}
    \label{fig:en_radius}
\end{figure}

Astrophysical X-ray photons will always arrive at the detector on the entrance window side, having passed through the telescope optics. Subsequently, they will be absorbed after passing through a depth of silicon drawn from an exponential distribution, characterised by an attenuation length that varies as a function of photon energy. Low energy photons will tend to be absorbed in very shallow layers of the detector medium where they liberate the cloud of electrons that produces the signal. In a back-illuminated device, the initial layers of silicon are furthest from the readout gates, thus the electron clouds will diffuse to larger radii by the time they reach the gates. On the contrary, higher energy photons are increasingly likely to penetrate further into the silicon before being absorbed, resulting in a charge cloud being liberated closer to the readout gates, resulting in smaller cloud radii. We simulate the absorption and subsequent charge cloud diffusion for a sample of X-ray photons, and the distribution of cloud radii as a function of photon energy is shown in Fig.~\ref{fig:en_radius:xray}.

Cosmic ray-induced background events, on the other hand, may enter the detector from any direction. The valid background events that are not rejected using traditional filtering methods (\textit{i.e.} single, double, triple and quadruple pixel events in valid patterns, falling below the energy threshold for rejection as a charged particle event) will predominantly result from secondary photons or low energy electrons or protons that result from interactions between the high-energy cosmic ray particles and other parts of the spacecraft. For each of the valid events produced in the \textsc{geant4} simulation, we calculate the total energy deposition and the cloud radius resulting from the mean energy deposition depth (weighted by the energy deposited at each step) to produce a comparable event energy \textit{vs.} cloud radius distribution, shown in Fig.~\ref{fig:en_radius:gcr}.

While astrophysical X-ray events follow a single track in their event energy \textit{vs.} cloud size distribution, with low energy events (absorbed further from the readout gates) producing a relatively narrow distribution of larger cloud radii, and higher energy events producing a broader distribution spanning from large to small cloud radii, the (unrejected) cosmic ray-induced events produce two tracks in their distribution. The upper track, centered at larger cloud radii for low energy events, spreading to smaller clouds for the higher energy events, results largely from the secondary photons that enter the detector from the entrance window side. A mirrored version of this track, centred at small cloud radii at low energies, expanding to larger radii at higher energies, is produced by approximately 50 per cent of the secondary particles that enter the detector from the opposite side (depending upon the shielding on around the detector and the relative fraction of background events entering from each side). The lower energy photons are absorbed in a relatively shallow depth of silicon. In this case, however, this means they are absorbed close to the gates, giving the clouds little opportunity to diffuse before readout. Additional events are produced with intermediate radii, notably in a line spanning the full range of cloud radii around 1.1\keV\ resulting from electron and positron events, and an island centred at 6\um\ cloud radii above $8\sim 10$\keV, which results from protons in addition to secondary photons produced as protons traverse the silicon.

If measurements can be made of the charge cloud radius in addition to the event energy, it may be possible to filter up to half of the `valid' background events, by rejecting those events whose cloud radii lie outside the distribution expected for genuine photon events of a given energy. Such characteristics of the raw event data may either be learned by AI algorithms, or a separate cut could be performed on the reconstructed properties of each event. This is possible using the Gaussian fitting event reconstruction scheme outlined above, fitting both the total event energy via number of electrons, $N_\mathrm{e}$, and the one-sigma radius of the Gaussian charge cloud, $\sigma$.

\begin{figure}
    \centering
    \subfigure[] {
    \label{fig:cloud_size:wfi}
    \includegraphics[width=0.3\linewidth]{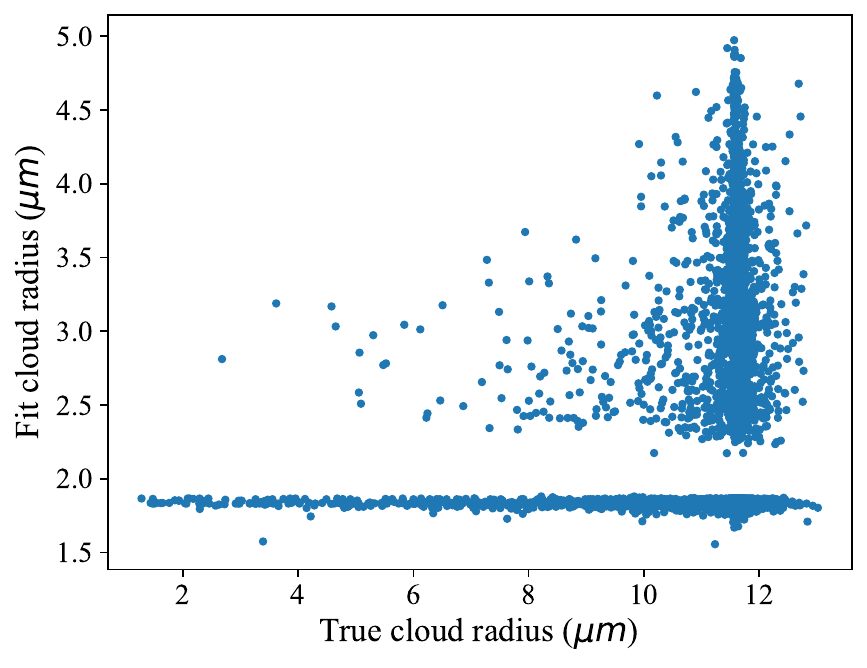}
    }
    \subfigure[] {
    \label{fig:cloud_size:axis}
    \includegraphics[width=0.3\linewidth]{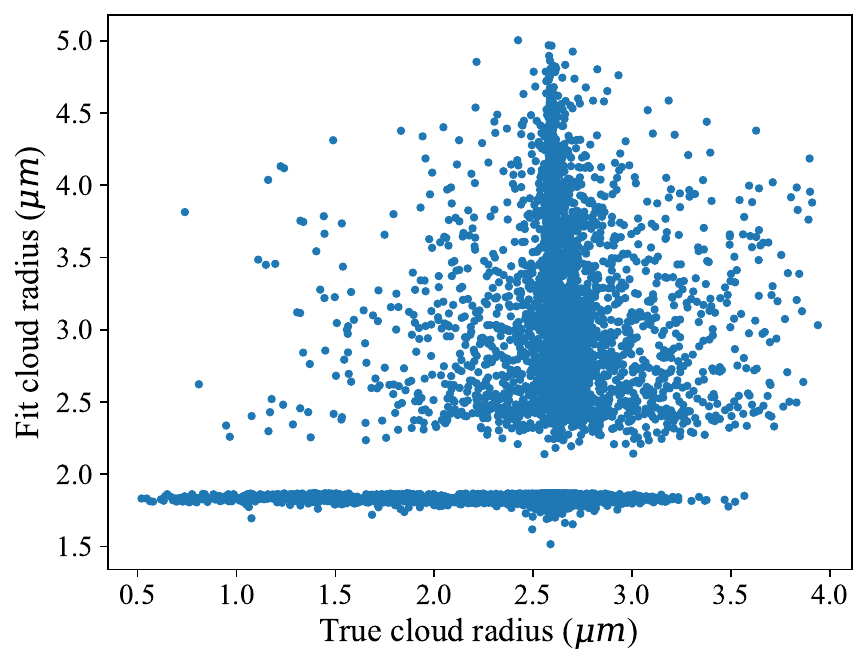}
    }
    \subfigure[] {
    \label{fig:cloud_size:difftest}
    \includegraphics[width=0.3\linewidth]{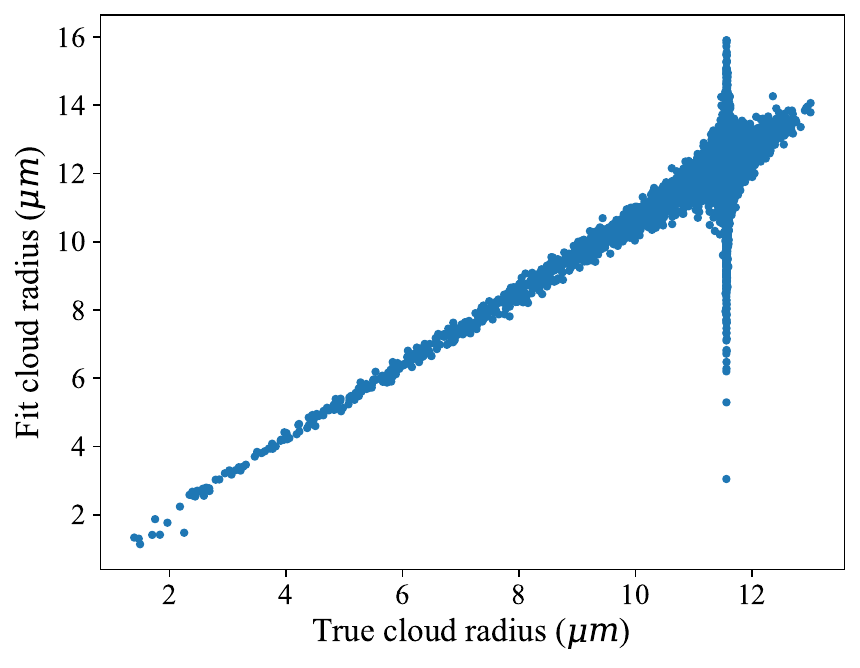}
    }
    \caption{The true electron cloud radius (defined by the standard deviation of the two-dimensional Gaussian electron cloud) at the readout plane within the detector \textit{vs.} the measured cloud size from the best-fitting $\sigma$ parameter in the Gaussian reconstruction method, for X-ray photon events in \subref{fig:cloud_size:wfi} an \textit{Athena WFI}-like device, with 130\um\ width by 450\um\ depth pixels, \subref{fig:cloud_size:axis} an \textit{AXIS} CCD-like device with 24\um\ width by 100\um\ depth pixels, and \subref{fig:cloud_size:difftest} a high-diffusion device with small, deep pixels measuring $10\times 100$\um. It is not possible to constrain the charge cloud size by fitting the pixel values in the \textit{Athena WFI} or \textit{AXIS}-like device, however when the charge cloud is able to diffuse across enough pixels to sufficiently resolve the cloud, such a measurement can be made using the Gaussian reconstruction method.}
    \label{fig:cloud_size}
\end{figure}

Successful measurement of the charge cloud radius requires the extent of the Gaussian charge cloud to be sampled to a sufficient degree by the pixels. We find that for the pixel size and depth planned for the \textit{Athena WFI} and \textit{AXIS} CCDs is not sufficient to place meaningful constraints on the cloud size (Fig.\ref{fig:cloud_size}). Even though the pixels of the \textit{AXIS} CCDs are smaller than those of the \textit{Athena WFI}, at 24\um\ pitch, compared with 130\um, the reduction in pixel depth from 450\um\ to 100\um\ results in the maximum diffused cloud radii dropping from around 12\um\ to around 3\um, thus are still under-sampled.  We note, however, that even when the cloud size is not constrained, accurate constraints can still be placed on the event energy by the Gaussian reconstruction method, as shown above.

When the pixel size is decreased relative to the depth such that the charge clouds are able to diffuse significantly into the surrounding pixels, we find that it is possible to measure the cloud size by fitting the Gaussian model to the measured pixel values. Fig.~\ref{fig:cloud_size:difftest} shows the measured cloud sizes for a 100\um\ thickness device for which the size of the pixels has been reduced to 19\um. Due to the larger cloud radii, it is necessary to fit the events in $11\times 11$ pixel rather than $5\times 5$ pixel regions. The additional diffusion, however, comes at the expense of a larger fraction of the event energy being lost into pixels on the edge of the event, leading to severe inaccuracies in the measurement of event energies using the traditional technique of summing pixel values that exceed the threshold. The Gaussian reconstruction method, however, is able to still accurately measure event energies, even in small, deep pixel, high diffusion devices.

\section{Conclusions}
Augmenting next-generation astronomical X-ray imaging detectors, including the CCD and DEPFET detectors that will underpin many next-generation flagship and probe-class X-ray missions, with artificial intelligence (AI) algorithms to analyse the raw frame-by-frame data holds the potential to enhance the sensitivity of these devices and reduce the instrumental background signal that is induced by cosmic ray particles interacting with the spacecraft and detector.

AI algorithms based on convolutional neural networks (CNNs) are now able to reduce the unrejected particle background signal by up to 41.5 per cent in the 0.3-10\keV\ band compared to traditional filtering methods that are the current state-of-the-art, and by 45.1 per cent over the 4-8\keV\ band, containing the iron K line that is of particular interest for a wide range of science cases. Just  1.8 per cent of the genuine X-ray events are lost when the AI filtering algorithm is employed.

The latest-generation AI algorithm employs a CNN to classify the frame and identify features in the image that distinguish cosmic ray-induced events from valid X-ray events, and then compute class activation maps (CAMs) that show the locations within the image containing the features that identify either cosmic ray or X-ray events. A random forest classifier is then used to compute the final classification of each event that is detected within the frame. The reduction in unrejected background comes from the ability of the AI algorithm to identify a cosmic ray-indusced event not only by the pattern of illuminated pixels, but by incorporating the context of other events detected within the same frame. By learning the characteristic spatial correlations that arise between secondary particles arising from the same primary cosmic ray particle, the algorithm is able to identify `valid' background events that are not rejected by traditional methods.

An alternative method for reconstructing detected photon events, based on fitting a Gaussian model of the charge cloud produced by the photon within the detector to the measured pixel values, is able to address shortcomings in traditional event reconstruction methods. Traditional methods can be prone to losing a significant fraction of the energy of photon events below 1\keV\ as the electrons carrying some of the energy diffuse into pixels that do not reach the split threshold. Noise-free simulations show that the Gaussian reconstruction method has the potential to enable accurate energy reconstruction and improved spectral resolution at the lowest photon energies that will be important for high-redshift science cases targeted by next-generation observatories. Moreover, it may be possible to obtain additional information that can distinguish genuine X-ray from particle-induced background events by measuring not just the energy but also the charge cloud radius, where the cloud can be sampled at sufficiently high resolution in small pixel devices.

\acknowledgments 
 
This work has been supported by the the NASA \textit{Astrophysics Research and Analysis} (APRA) program under grant number  80NSSC22K0342 as well as the US \textit{Athena Wide Field Imager} Instrument Consortium under NASA grant NNX17AB07G.

\bibliography{aibkg} 
\bibliographystyle{spiebib} 

\end{document}